# Transparent and heat-insulation bionic hydrogel-based smart window system for long-term cooling and waste heat collection

Qianwang Ye[1†], Hanqing Dai[1†*], Yukun Yan[1], Liwei Wang[1], Xinlin Du[1], Yimeng Wang[2], Zhile Han[3*], Wanlu Zhang[1*], Ruiqian Guo[1*]

[1]*College of Intelligent Robotics and Advanced Manufacturing, Fudan University, Shanghai 200433, China.*
[2]*School of Science and Engineering, University of Dundee, DD1 4HN, Dundee, United Kingdom.*
[3]*Suzhou Institute of Biomedical Engineering and Technology, Chinese Academy of Sciences, Suzhou, Jiangsu 215163, China.*

[†]These two authors contributed equally to this work.

Corresponding author email: daihq@fudan.edu.cn; hanzl@sibet.ac.cn; fdwlzhang@fudan.edu.cn; rqguo@fudan.edu.cn;

**Abstract:** With the energy crisis and climate warming, the position of a new generation of smart windows is becoming increasingly important, and materials or systems that can have high blocking of near-infrared (NIR) and ultraviolet (UV) and high transmittance of visible light (VIS) are needed. Currently, it is difficult for smart heat-insulation materials to achieve high transmittance of VIS, good UV isolation, outstanding cooling and thermal insulation, and excellent waste heat collection. Here, we design a novel composite hydrogel to achieve an average 92% VIS transmittance, efficient UV absorption (>75%), 11 °C of thermal insulation, and sensing properties. Interestingly, we designed a transparent heat insulation system with this composite hydrogel to obtain about 22 °C of the record-breaking insulation performance for 168 hours, waste heat collection and reutilization, and temperature sensing. Our findings provide new ideas and possibilities for designing transparent and heat-insulation smart window systems.

## 1. Introduction

With the intensification of global warming and the increasing frequency of high-temperature warnings worldwide, the challenges of environmental degradation and the energy crisis have become more severe. One of the critical issues in sustainable development is how to minimize energy loss[1, 2]. Among various energy consumption sectors, cooling during high-temperature seasons demands substantial resources, drawing significant attention to the development of effective heat insulation materials. In urban and residential areas, transparent glass is widely used in buildings, making it a key factor in both energy efficiency and living comfort[3, 4]. Enhancing the thermal insulation properties of glass not only reduces energy consumption but also improves indoor environmental quality. Under this background, transparent thermal insulation materials have emerged as a vital component in smart glass technology, offering remarkable applications in building energy conservation[5-7], electronic devices[8, 9], healthcare[10], environmental protection[11], and broader sustainable development efforts[12].

Currently, commonly used transparent thermal insulation materials in buildings primarily include vacuum or inert gas-filled glass (e.g., argon-filled glass)[13, 14] and coatings/films based on metals or metal oxides (e.g., single/double silver-based films, indium tin oxide (ITO), and antimony tin oxide (ATO))[5, 15, 16], and some researches use other materials, including porous materials[17, 18], multilayer materials[19, 20], photonic crystals[21], and biomimetic materials[22, 23]. However, these materials involve a trade-off between thermal insulation performance and optical transparency. These materials have some shortcomings. Such as metal or metal oxide coatings effectively reduce heat transfer but may lower visible light transmittance, exhibit angle-dependent optical properties, and suffer from durability issues of heat insulation over time[24]. Moreover, transparent thermal insulation materials face challenges related to complex manufacturing processes that demand high-precision control and specialized materials, resulting in high production costs and limiting large-scale application[25]. Furthermore, prolonged exposure to sunlight and ultraviolet radiation accelerates material aging, leading to performance degradation, yellowing, and reduced durability[26]. Importantly, these materials often lack stimulus-responsive features, limiting their intelligent functionality[27]. Consequently, to achieve an optimal balance between smart energy efficiency and visual comfort, developing next-generation thermal insulation materials with enhanced transparency, improved durability, and high intelligence has great application potential.

Gel-based materials have garnered increasing attention as effective thermal insulation solutions[28, 29]. A common approach involves dissolving metal-organic compounds in water or organic solvents and molding them under controlled temperatures[30]. Among gel-based materials, aerogels stand out for their exceptional thermal insulation performance. By replacing the liquid phase in gels with air, aerogels form nanostructures characterized by low solid thermal conductivity, effective radiant heat insulation, and highly porous structures[31]. These properties have led to their widespread use in advanced thermal insulation applications. Notably, the incorporation of reflective materials such as silica significantly reduces light transmittance, often dropping to as low as 50%[32]. Currently, one of reported studies developed a bio-based photoluminescent aerogel that achieved a record cooling effect of 16 °C[33]. However, this opaque aerogel still has low cooling effect, poor stability, the lack of VIS transmittance, waste of heat, and inherent brittleness[29].

To address, we imitated the unique structure of the green-leaf plant to design a novel bionic device with thermal insulation and a smart sensing layer as well as a cooling and collection-reuse heat layer. This device utilizes a carbon quantum dot (namely, CPO-CDs)-composite polyacrylamide (PAM)/polyacrylic acid (PAA)/betaine (namely, CPPB) hydrogel as a novel thermal insulation and smart sensing layer, and the water circulation layer is used for system cooling and collection-reuse heat. Our CPPB hydrogel has high VIS transmittance (average 92%), low UV transmittance (<20%), and low NIR transmittance (<30%). Under the premise of high transparency, the CPPB hydrogel achieved 10-14 °C cooling and a high UV barrier (>75%) in a vacuum environment. Furthermore, we designed the smart window system (namely, the CPW system) with 21.6-21.9 °C of the insulation performance for 168 hours, and we obtained a more efficient energy utilization system that can use waste heat to heat the water circulation. Our study provides feasible solutions for the design of a new generation of smart windows and smart homes in the future.

## 2. Results and Discussion

### Design, Characteristics, and Properties of CPPB Hydrogel

As we all know, green-leaf plants dissipated their heat through transpiration to avoid death due to overheating (**Figure 1a**). In the high-temperature environment, water is transported upward from the roots and then released through the stomata of the leaves. When the water changes from liquid to gas, it absorbs a lot of heat, which reduces the temperature of the blade. Inspired by this, we designed a thermal insulation material and a new bionic system (**Figure 1b**) that mimics the unique structure of green-leaf plants, with thermal insulation and intelligent sensing layers, as well as cooling and collection-reuse of heat. Here, we prepared CPPB hydrogels by free radical polymerization, achieving mechanical stability, UV protection, anti-aging, heat insulation and smart sensing. In particular, the cooling system, which consists of circulating water and mechanical pumps, takes away the heat absorbed by the bionic device and uses solar panels to provide hot water for daily life with zero carbon (**Figure 1c**).

It is acknowledged that NIR (760 < λ < 2400 nm), which makes up about 50% of sunlight, carries significant heat. UV (200 < λ < 400 nm), accounting for 7% of sunlight and UV can more effectively promote most photochemical degradation mechanisms, such as photochemical cleavage, photo-oxidation and discoloration[34, 35]. The ideal smart glass material should exhibit high shielding for UV and NIR wavelengths, high transmittance for the VIS range, and high emissivity in the atmospheric window. To achieve this, we selected hydrogel materials with high transparency in VIS, and previous studies have proved that the amide bond (CO−NH) has certain thermal and light stability[36], so we used polyacrylamide as the main framework of hydrogel. We prepared CPPB hydrogel by free radical polymerization. Their high-water content provides excellent VIS transmittance while offering a solid foundation for blocking infrared wavelengths. We tested the absorption and transmittance rates of the CPPB hydrogel from 200 to 2500 nm, demonstrating the material's performance across this spectrum (**Figure 1d**). The results show that CPPB hydrogel exhibits significant absorption in both the UV (>75%) and NIR (>70%) regions while maintaining an average VIS transmittance of 92%. The high infrared absorption of CPPB is primarily attributed to the intrinsic vibrations of key chemical bonds, including the stretching vibration (3.125 μm) and bending vibration (6.25 μm) of the hydrogen bond O−H from deionized water and including C−H and CO−NH, which we tested in the hydrogel after freeze-drying to remove water, (**Figure 1e**). It has already been shown that the intrinsic bond vibrations within the atmospheric transparency window affect the emissivity of materials[37]. The chemical bonds, which vibrate at wavelengths from 8 to 13 μm and from 16 to 24 μm, occur precisely within the transparent window of the atmosphere, allowing the hydrogel to radiate energy efficiently.

In order to further strengthen the external radiation effect of chemical bonds in the atmospheric transparency window, we used F127 to introduce ether bonds (C−O−C) and tried different contents to influence the thermal insulation performance of CPPB hydrogel (**Figure 1f**). And we used an infrared heat source lamp to release heat to the materials and measured the temperature changes 10 cm below different materials with a temperature sensor. We confirm that the introduction of F127 brought about ether bonds (**Figure S1**), and under the action of F127, the results indicate that insulation performance improves slightly with increasing F127 content, reaching its optimal effect at a mass fraction of 5%. At this concentration, the ether bonds in F127 effectively vibrate and absorb energy under 1500 nm infrared irradiation. However, when the mass fraction increased to 25%, the hydrogel became no longer transparent, leading to a significant reduction in VIS transmittance. Therefore, a mass fraction of 5% was selected as the optimal composition.

The thermal insulation properties of CPPB hydrogel were further evaluated under different water content levels over a two-hour period. The results demonstrate that CPPB hydrogel achieves the best insulation effect when the water content is maintained between 50% and 70% (**Figure 1g**). This proves that the thermal insulation properties of CPPB hydrogel are still caused by the chemical bonds of the hydrogel, rather than the hydrogen bonds in water. When the water content ranged from 0% to 40%, an increase in water content led to greater stretching of chemical bonds and a higher number of hydrogen bonds. This effectively delays heat transfer and penetration, thereby enhancing thermal insulation. Simultaneously, as water content increased, VIS transmittance of the hydrogel also improved (**Figure S2**). However, when the water content exceeded 60%–78%, the CPPB hydrogel network underwent continuous expansion, leading to a loss of structural integrity.

This resulted in reduced toughness and mechanical stability, making the material less suitable for practical applications in long-term scenarios.

To further evaluate the thermal insulation performance of CPPB hydrogel, we conducted an experiment in which the material was sealed between two thin glass plates (1 mm thickness). A control group was established using two thin glass plates with a vacuum interlayer. An infrared heat source (wavelength range: 760 nm–5000 nm) was applied, and the environmental temperature 10 cm below the experimental setup was recorded (**Figure 1h**). The results indicate that CPPB hydrogel maintains effective thermal insulation for up to two hours. During the initial heating phase, temperatures in both the control and experimental groups increased (the experimental chamber was fully sealed, measuring 10 cm × 10 cm × 10 cm, with light shielding and heat insulation on all sides to minimize heat conduction and convection effects). However, once the heat source stabilized at 85 °C with an infrared radiation intensity of 450 W/m², a significant difference emerged. The CPPB-containing setup maintained a stable temperature difference of 11 °C (ranging from 10 °C to 14 °C) compared to the control group, and this effect was sustained for up to two hours. These findings suggest that CPPB hydrogel exhibits excellent long-term thermal stability, demonstrating strong potential for practical applications. And finally prove the rationality of our above work and achieve the expected goal.

To enhance UV-blocking capabilities, CPO-CDs were synthesized and analyzed using UV-VIS's absorption spectroscopy (**Figure S3**). The absorption spectrum reveals three strong absorption bands, including a broad band in the 200–300 nm range and two distinct peaks at 302 nm and 361 nm. The high absorption in the 200–300 nm range is attributed to the presence of multiple conjugated ring structures and functional groups in the carbon shell of the CPO-CDs. Ortho-phenylenediamine (oPD), a commonly used nitrogen source in CPO-CDs synthesis, plays a crucial role in modifying the absorption properties. In the presence of citric acid (CA) and polyethyleneimine (PEI), oPD undergoes hydrolysis, leading to the formation of pyrrolidine-N[38, 39]. This process, coupled with hydrothermal carbonization, increases the number of benzene rings, resulting in a denser carbon core. As a consequence, more conjugated structures are formed, generating new absorption bands while preserving the original absorption characteristics. The composition and solid-state UV absorption peak at 307 nm are directly linked to the aromatic sp² structures within the CPO-CDs. The presence of multiple conjugated ring structures and functional groups in the carbon shell further contributes to a strong absorption band in the 200–250 nm range.

Comparing the UV absorption spectra of CPPB and the hydrogel without CPO-CDs added (namely, PPB)

(**Figure 1i**), we observe that the synthesized CD-hydrogel composite exhibits a pronounced absorption band in the 250–300 nm range, along with a new absorption peak near 351 nm. Notably, this peak shifts from the original 361 nm (observed in CPO-CDs alone) to 350 nm, indicating an interaction between the CPO-CDs and the hydrogel matrix. This spectral shift confirms that the incorporation of CPO-CDs into the hydrogel is not a mere physical mixing but involves strong interfacial interactions. Moreover, the composite hydrogel material demonstrates broad-spectrum UV shielding under sunlight exposure, effectively providing UV protection and anti-aging properties. This enhancement significantly expands the potential applications of the CPPB hydrogel, particularly in advanced glass interlayer materials and other functional coatings. Based on the experimental data, we observed that our CPPB hydrogel exhibits a high blocking efficiency in UV (>75%). We used commercially available UV test paper to test the UV absorbance properties of our CPPB hydrogel in outdoor sunlight, and the results showed that its UV blocking effect was obvious (**Figure S4**). This property helps mitigate the degradation effect of UV from sunlight on indoor furnishings. Interestingly, recent researchers have discovered that photoluminescence can convert UV into VIS, which enhances the radiation of VIS and helps dissipate a certain amount of heat[33].

## Design and Properties of CPW System

Currently, some materials, while insulating heat transfer, experience an increase in their temperature, which poses significant challenges to the material's thermal conductivity and stability[40]. There is currently no suitable method for managing the generated waste heat. The most common approach involves allowing the material to radiate thermal energy into the cooler air through the mid-infrared atmospheric windows(8-13 and 16-24 μm)[41,42]. In response, we propose a CPW system that utilizes water circulation to collect the waste heat, thereby enabling efficient energy utilization.

To further improve the thermal insulation performance and heat collection capabilities of the CPPB hydrogel, we developed a bionic system inspired by the transpiration process found in green plants. This CPW system is powered by solar panels and uses circulating cold water to transfer the heat absorbed by the CPPB hydrogel into domestic water. When the circulating water reaches a predefined high temperature, it is then directed to a domestic water heater. We set up a CPW system with five layers of structure, using three layers of quartz glass with a thickness of 1 mm to clamp the CPPB hydrogel layer and condensate layer with a thickness of 2 mm respectively. As shown in **Figure 2a**, it is the physical diagram of the simplified CPW system device. The condensate layer is connected to a pool with two access pipes, and a pump is used to power the water circulation. Using the height difference, the low water inlet is fed cold water, and the high-water outlet is exported to ensure that the water layer is full. When the CPPB hydrogel absorbs heat and then transfers it to the condensate layer, the heat is collected through the condensate layer for heating domestic water and realizing the secondary utilization of waste heat. In addition, we used a solar panel to meet the energy required by the pump, which truly realizes energy saving and emission reduction.

The most important structure of this CPW system is still the five-layer structure. To ensure that there is still excellent VIS transmittance, we tested the refractive index of each layer and calculated its theoretical transmittance (**Figure 2b**). First, we measured and found that the refractive index of our CPPB hydrogel layer is $n_{CPPB} = 1.4$, the refractive index of the glass is $n_{glass} = 1.5$, and the refractive index of the condensate water layer is $n_{water} = 1.33$. The reflectance R is calculated by Fresnel's formula (**Eq. 1**), where $n_1$ and $n_2$ are the refractive indices of the two media respectively. The transmittance T is calculated by **Eq. 2**. The transmittance between each layer can be obtained as: $T_{glass-CPPB} = 99.88\%$, $T_{CPPB-glass} = 99.88\%$, $T_{glass-water} = 97\%$, $T_{water-glass} = 97\%$ and according to the **Eq. 3**, the comprehensive transmittance $T_{total} = 93.86\%$. This shows us that our multilayer structure still retains a high VIS transmittance.

$$R = \left(\frac{n_1 - n_2}{n_1 + n_2}\right)^2 \tag{1}$$

$$T = 1 - R \tag{2}$$

$$T_{total} = \prod T \tag{3}$$

To further explore the optimal temperature of the system, we established a molecular dynamics model (**Figure S5**). Considering the issue of energy collecting efficiency, we calculated the volume changes of CPPB hydrogel at different temperatures (**Figure 2c**). It can be observed that, for the same number of molecules, the CPPB hydrogel exhibits the most compact molecular arrangement and the smallest volume at 40 °C, at which point the hydrogel has the highest density. Under same conditions, more heat can be absorbed for the same volume. Meanwhile, from the data, we can find that within the temperature range of 30-80 °C, the volume change of CPPB hydrogel is only 1%, which simultaneously proves that this material has good thermal stability and low thermal expansion. Additionally, we calculated the ratio of enthalpy change to temperature change for the CPPB hydrogel at different temperatures (**Figure 2d**), which demonstrates that the specific heat capacity of the material is maximized at 40 °C. Based on both the density and specific heat capacity considerations, we chose to maintain the CPW system at 40 °C for optimal heat collection efficiency.

As shown in **Figure 2e**, prior to switching the domestic water supply, we were able to consistently heat the water to 37 °C. This continuous cycle not only enhances the insulation performance but also facilitates the rapid supply of domestic water, resulting in a 20 °C reduction in temperature. Furthermore, by incorporating the temperature-sensing capabilities of the hydrogel, when the indoor temperature is lower than the outdoor temperature, the collected circulating hot water can be redirected back into the biomimetic CPW system, which can prevent fog formation that typically occurs when warm air contacts cold surfaces[43]. This bionic CPW system effectively maintains a comfortable indoor environment while supporting a zero-energy, sustainable lifestyle.

## Smart Sensing Integrated Design

To verify the potential of CPPB hydrogel as a future generation of smart windows, the sensing properties are essential. At present, some research on the sensing of smart window materials mainly reflects the temperature sensing and photochromic parts[44]. Because cationic groups and anionic groups are brought by the zwitterion betaine in the CPPB hydrogel, it has a certain electrical conductivity. Since electrical conductivity is key to enabling a new generation of smart material sensing and monitoring, it is also necessary to test their electrical conductivity[45-48]. Thus, we tested the conductivity of CPPB hydrogel (**Figure S6**). Conductivity measurements determined that the material exhibits a conductivity of 5 S/m, a relatively high value within the field of smart home technology, demonstrating its potential for integration into smart systems.

As the interlayer of a new generation of smart glass, it needs certain deformation resistance and compression resistance. In order to verify the specific structure of CPPB hydrogel, scanning electron microscopy (SEM) was used to further examine the morphological characteristics of the hydrogel (**Figure 3a**). The SEM images show a well-defined layered structure with internal voids, each layer being about 20 μm thick. This confirms the successful formation of three-dimensional networks, which are essential for the multifunctional properties of hydrogel[49]. Previous studies have shown that acrylic acid and betaine can form an ionic elastomer with excellent mechanical properties[50], but this material is soluble in water, and we used acrylamide as the main skeleton on its basis to achieve structural stability. Then we evaluated the mechanical behavior of CPPB hydrogel. **Figure 3b** illustrates its stress−strain curves, indicating that the hydrogel possesses a certain degree of elasticity. This CPPB hydrogel is capable of elongating to twice its original length, demonstrating excellent resistance to deformation. This flexibility is a crucial advantage for applications requiring mechanical durability and adaptability.

To assess the CPPB hydrogel's ability to withstand external forces, we tested its compressive performance[51, 52] (**Figure 3c**). A key parameter for evaluating strain sensors is the gauge factor (GF), which quantifies the sensitivity of the material to applied strain. It is defined as **Eq. 4**, where ΔR represents the change in resistance, $R_0$ is the initial resistance, and ε is the applied strain. Experimental results indicate that CPPB hydrogel exhibits a GF of 0.18 in the 0-100 kPa pressure range, suggesting that the insulating interlayer maintains stability under pressure. This implies that CPPB hydrogel can withstand pressures up to 100 kPa without compromising its sensing performance.

To further explore the multifunctionality of CPPB hydrogel, we examined its temperature perception capabilities (**Figure 3d**). Previous studies have confirmed that the conductivity of CPPB hydrogel is positively correlated with temperature, as increased temperature promotes ionic dissociation. Leveraging this property, CPPB hydrogel can be utilized as a thermal sensor to detect temperature fluctuations in real time. To quantify its temperature sensitivity, we introduced the temperature coefficient of resistance (TCR)[53, 54], defined as **Eq. 5**, where $R_0$ is the initial resistance, ΔR is the relative resistance change, and ΔT represents the temperature variation. For better accuracy, CPPB hydrogel's response region was divided into two segments within the temperature range of 30-100 °C. As shown in **Figure 3e**, the TCR value was determined to be -0.61 %/°C in the range of 30-70 °C and -0.42%/°C in the range of 70-100 °C. Furthermore, the CPPB hydrogel shows approximately the same warming and cooling process over a continuous cycle of 30-50 °C (**Figure 3f**), indicating that the CPPB hydrogel has good cycling stability to temperature. This flexible and responsive temperature sensing capability allows CPPB hydrogel to dynamically detect external temperature variations and translate them into real-time resistance signals. Such functionality enables interactive temperature monitoring, making CPPB hydrogel highly promising for applications in smart home environments and intelligent ecological systems.

$$\text{GF} = \frac{\Delta R/R_0}{\varepsilon} \tag{4}$$

$$\text{TCR} = \frac{\Delta R/R_0}{\Delta T} \tag{5}$$

## Application and Energy Saving

To assess the practical performance of the CPPB hydrogel, we conducted an outdoor field test on October 11, 2024, at Fudan University's Jiangwan Campus, located at 121.5°E and 31.3°N. The test took place on the roof of Building 2, where we evaluated the thermal insulation properties of the CPPB hydrogel (**Figure 4a**). The temperature on that day was between 18 and 25 °C with an RH of 93%. This experiment involved comparing the thermal insulation effects of the CPPB hydrogel to those of a control group under direct sunlight. On the day of testing, the weather was clear with an ambient temperature of 25 °C. During the period of concentrated solar radiation (from 14:00 to 16:00), the data revealed that the transparent insulation material achieved a maximum thermal insulation difference of 10.8 °C. To further prove the feasibility and stability of the CPW system, we used a solar simulator for a one-week exposure test (**Figure 4b**). The heat source is controlled at 83 °C, and for up to 168 hours, the CPW system (38.1-38.4 °C) can achieve a stable cooling of 21.6-21.9 °C compared with the system without (60 °C), and the water temperature is maintained at 32.7-33.4 °C, which also reflects the long-term stability of the system and can cope with the high-temperature climate of longer daylight exposure. Compared to some of the different materials used by other researchers (**Figure 4c**)[5, 12, 15, 28, 31, 33, 55-65], our CPPB hydrogel and CPW system demonstrated notable advantages, both in terms of VIS transmittance and thermal insulation effect.

At the same time, to further evaluate the practical application potential of CPW system, we simulated the cooling energy savings in 30 different cities around the world using a typical mid-rise apartment model. Their geographical locations are shown in **Figure 4d**. The selected cities include tropical, subtropical and temperate climates. The coordinates and climates are shown in the table (*detailed in* **Table S1**). We compared our insulation CPW system with vacuum glass, which is more commonly used today, and calculated the annual cooling energy difference between the two to demonstrate the practicality of our work. In **Figure 4e**, we found that the annual cooling energy savings were even more pronounced in the tropics, especially in New Delhi and Abu Dhabi, where annual cooling energy savings reached 569.1 MJ/m$^2$ and 512.5 MJ/m$^2$. In subtropical monsoon climates and Mediterranean climates, such as Hong Kong and Barcelona, the annual cooling energy savings are still good, reaching 333.9 MJ/m$^2$ and 363.3 MJ/m$^2$, while in temperate regions, the annual cooling energy savings are much lower. Moscow and Amsterdam, for example, reached 156.0 MJ/m$^2$ and 145.1 MJ/m$^2$, but this trend is also compounded by reality[41]. This is because there is more solar insolation in the tropics than in temperate regions, so we calculated the percentage reduction in energy consumption (**Figure 4f**). And we found that its trend is opposite to the trend of annual cooling energy savings. Temperate regions such as Moscow (34.2%) and Berlin (31.3%) experienced higher percentage reductions than tropical regions such as Mexico City (13.2%) and Singapore (5.4%). Compared to the latter, the former has shorter periods of direct sunlight, lower average temperatures, and less total cooling energy consumed per year, resulting in a lower percentage of energy consumed. The simulation results of the other 24 cities show that the annual cooling energy consumption reducing provided by the film are between 87.2 and 569.1 MJ/m$^2$ (Mexico City (310.6 MJ/m$^2$), Bangkok (205.1 MJ/m$^2$), Cairo (428.4 MJ/m$^2$), Janeiro (136.6 MJ/m$^2$), Havana (412.0 MJ/m$^2$), Bandar Seri

Begawan (194.1 MJ/m$^2$), Singapore (197.2 MJ/m$^2$), Istanbul (293.9 MJ/m$^2$), Los Angeles (461.1 MJ/m$^2$), Shanghai (213.1 MJ/m$^2$), Inchon (148.2 MJ/m$^2$), Tokyo (223.9 MJ/m$^2$), Buenos Aires (100.6 MJ/m$^2$), Sydney (123.4 MJ/m$^2$), Cape Town (115.1 MJ/m$^2$), Madrid (321.8 MJ/m$^2$), Santiago (93.4 MJ/m$^2$), Melbourne (87.2 MJ/m$^2$), Vancouver (209.2 MJ/m$^2$), London (192.0 MJ/m$^2$), Geneva (196.4 MJ/m$^2$), Berlin (203.9 MJ/m$^2$), Beijing (191.1 MJ/m$^2$), New York (241.7 MJ/m$^2$)). The above theoretical simulation results show that our thermal insulation CPW system possesses exceptional cooling properties worldwide, which will help demonstrate its application and development capabilities and offer a potential solution for applications in energy-efficient and sustainable building design.

## 3. Conclusions

In summary, this study proposes a transparent and heat-insulation bionic hydrogel and a system to address the growing challenges of energy conservation and environmental sustainability. CPPB hydrogel has high VIS transmittance (average 92%), low UV transmittance (<20%) and low NIR transmittance (<30%). In addition, CPW system we designed includes a circulating water layer powered by solar panels, which effectively collects solar radiation heat and reuses it in domestic hot water, achieving a 168-hour temperature reduction of 21.6-21.9 °C, while providing temperature sensing. Combining sustainable high-efficiency heat insulation, UV protection and smart sensing, this new material and system design offers a promising solution for future intelligent building and energy-efficient building design, contributing to the sustainable development of the building industry and reducing energy consumption.

# Methods

## Materials and Reagents.

Raw materials comprise acrylamide (Am, 99%), acrylic acid (AA, 98%), N, N '-methylenebis(acrylamide) (Bis), ammonium persulfate (APS, AR), N, N, N', N'-tetraethylethylethylenediamine (TEMED, ≥99%), Pluronic F127 (F127, 97%), betaine (≥98%), citric acid (CA, ≥99.5%), Polyethylenimine (PEI, MW 1800, 99%), o-phenylenediamine (oPD, AR) and deionized water. All reagents were purchased from Aladdin Chemicals Limited.

## Preparation of CPO-CDs.

The CPO-CDs were prepared by using CA, PEI and oPD. First, 1 g of CA, 0.4 g of PEI and 0.4 g of oPD were wholly dissolved in 20 mL of deionized water with ultrasonic treatment for 30 min. Then the obtained solution was added to the Teflon-lined autoclave and heated at 180 °C for 8 h. After cooling to room temperature, the yellow solution was obtained by filtering the macromolecular particles with a 0.22 μm membrane filter. The solution was then redialed in a dialysis bag (500 Da) with deionized water for 24 h to remove the small molecule particles. Replace deionized water at 8 h intervals. Finally, the CPO-CDs powder was obtained by freeze-drying.

## Preparation of CPPB Hydrogel.

First, take 2 g of Am, add 1 mL of AA, 0.1 g of Pluronic F127, 2 g of betaine and dissolve in 13 mL of deionized water, stir until completely dissolved. Then add CPO-CDs powder 0.1g, 20 mg Bis, 200 mg APS, 400 μL TEMED into the mixed solution, dissolve and pour into a mold to complete the preparation.

## Design of CPW System.

We designed a CPW system based on the transparent insulation material CPPB hydrogel. Under the premise of CPPB hydrogel heat collection, the water layer is used for condensation for waste heat transfer and reuse. Three layers of quartz glass with a thickness of 1mm are used to clamp the hydrogel layer and the water layer with a thickness of 2 mm, and the four sides are sealed with high-temperature waterproof materials. At the same time, a pair of inlets and outlets of the water layer is set aside for the flow of condensate water. We use a pump to continuously circulate our condensate water into the system, while the waste can be used for other purposes. Since the pump does not require higher power, the energy requirements of the pump can be met by using solar panels. The self-operation of the system can be realized in the external environment with high temperature, and the hydrogel layer can transmit electrical signals to show the external temperature.

## Characterization of Morphology and Components.

The morphologies of the samples were obtained using a scanning electron microscope (SEM) at 5.0 kV with a ZEISS Sigma 300. Fourier transform infrared (FTIR) spectroscopy measurements were recorded with a Thermo Fisher Scientific Nicolet iS20.

## Transparency Measurement.

Ultraviolet−visible spectroscopy (UV-3600, Shimadzu) was utilized to measure the transmittance of ionic hydrogel (thickness 2 mm) with a wavelength range from 400 to 800 nm, and the measurement was operated at room temperature.

## Thermal Insulation Properties Characterization.

We designed a system, using a black box with a height of 10cm, the upper part of the box can be treated with light transmission. The dimensions of our box were selected as 10 cm × 10 cm × 10 cm, which were completely sealed and treated with light shielding and heat insulation on all sides. Try to eliminate the influence of heat transfer and heat convection.

## Mechanical Properties Characterization.

The samples with a size of 20 mm × 40 mm were tensed 1 cm along the long side, and the force values were recorded by a Suce SH-III digital push-60 pull meter. During rubbing and sticking with tape, a self-designed sticky hydrogel was designed and prepared for use as the tape. The tensile stress−strain behaviors of the prepared hydrogel were determined on an electronic universal testing machine (INSTRON 3367). The tensile tests were conducted at 25 °C, and the testing speed was 5 mm/min.

## Electrical Properties Characterization.

Measurement of Electrical Signals. Electrochemical impedance spectroscopy (EIS) was obtained with an electrochemical workstation (CHI660E) at room temperature. The ionic conductivity was calculated according to the equation:

$$\sigma = \frac{L}{R_b \cdot S}$$

where σ represents the ionic conductivity of the sample, L represents the thickness of the sample, $R_b$ represents the bulk resistance of the sample, and S represents the electrode contacting area. The other electrical signals were measured with a Keithley 2400 SourceMeter.

**Anti-UV Properties.**

Ultraviolet−visible spectroscopy (UV-3600, Shimadzu) was utilized to measure the absorption of ionic hydrogel (thickness 2 mm) with a wavelength range from 200 to 400 nm, and the measurement was operated at room temperature. And we used an ultraviolet sensor card to test its anti-UV properties in a normal outdoor environment.

**Thermal-Sensitive Performance Characterization.**

The ionic hydrogel samples (40 mm × 20 mm × 2 mm) were sealed in a waterproof plastic bag. The ends of the ionic hydrogel were connected to an electronic universal testing machine (INSTRON 3367) with copper wires and placed in a water bath. The temperature sensitivity test was performed by changing the temperature of the water bath, and the experimental data were recorded after the current fluctuation was stabilized to ensure the accuracy of the measurement.

## Acknowledgements

This work was supported by the National Key Research and Development Program of China (2024YFE0204600), National Natural Science Foundation of China (62305068 and 62074044), China Postdoctoral Science Foundation (2022M720747), Shanghai Post-doctoral Excellence Program (2021016), Shanghai Rising-Star program (22YF1402000).

**Author contributions:** H.D. and R.G. conceived and designed the project. Q.Y. Y.W., and Z.H. generated the data. Q.Y. and H.D. analyzed and interpreted the data. Q.Y. and H.D. wrote the manuscript. R.G. W.Z. H.Q. and Z.H. reviewed and revised this manuscript. R.G. W.Z. and Z.H. supervised the study. Y.Y., L.W. and X.D. assisted in the completion of this project. All authors edited and approved the manuscript.

**Competing interests:** Authors declare that they have no competing interests.

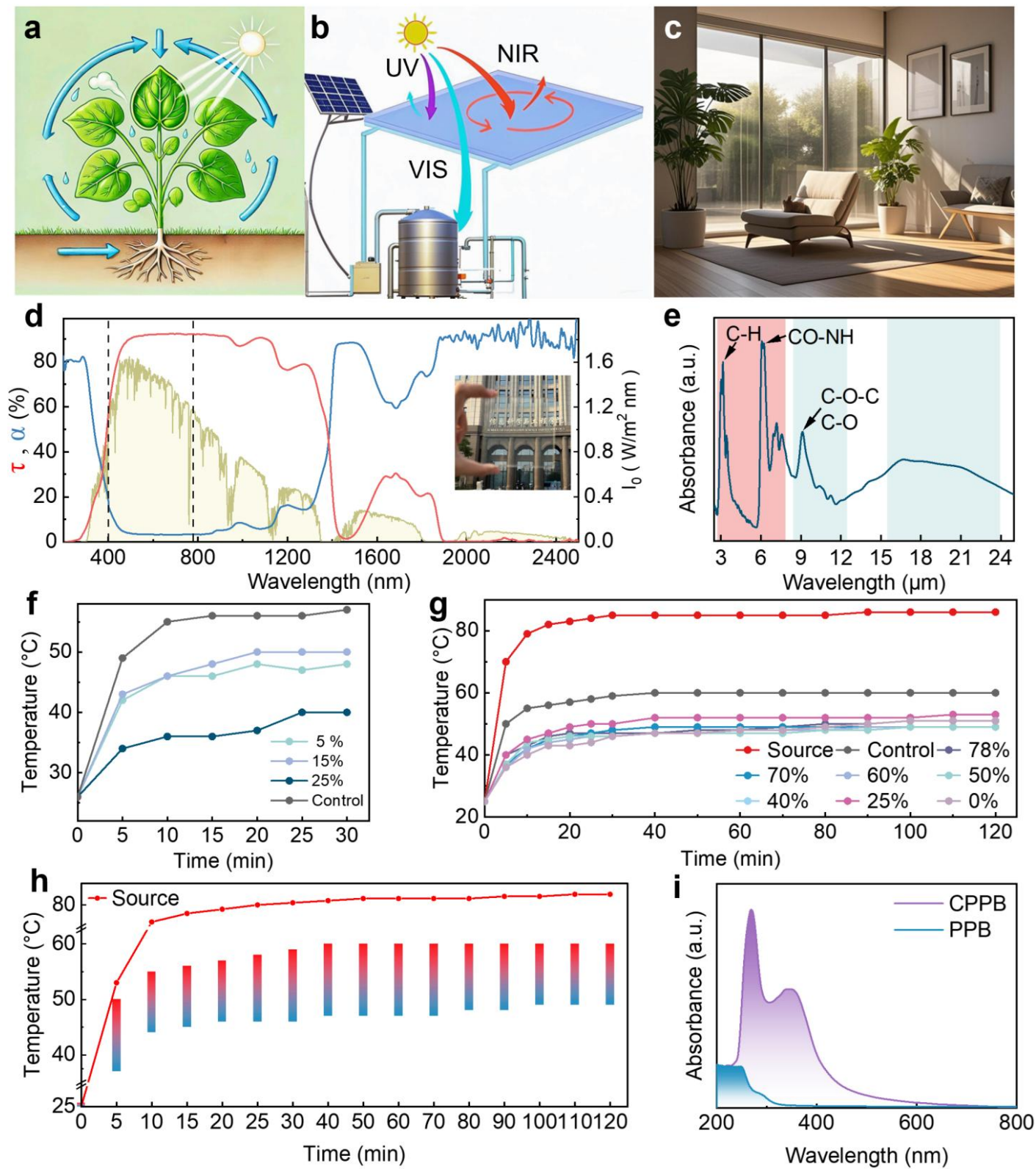

**Fig.1 Smart window design idea and optical properties of the transparent CPPB hydrogel. a**, Heat dissipation mechanism of green-leaf plants: water is collected and transmitted to the leaf surface through roots and xylem, and heat dissipation is achieved through transpiration and precipitation cycle under sunlight **b**, Working mechanism of the CPW system: Above is a layer of CPPB hydrogel structure that use CPO-CDs photoluminescence to absorb UV and convert it into VIS light emission, a large range of VIS light transmission and a large amount of heat collection and external radiation in NIR. Pumps and pipes are used to transfer water under CPPB hydrogel to collect heat and reuse it in domestic water. The system operates with a solar panel supply water pump, enabling zero-carbon system development. **c**, Vision of future smart home environment: Transparent insulated windows only pass high through VIS light, have certain smart sensing and achieve zero

carbon environment. **d**, Transmissivity τ (red line) and absorptivity α (blue line) spectrum of the metamaterial coating. The yellow spectrum represents the AM 1.5 Global solar reference spectrum: 49.9% of the solar irradiance is in the NIR, 45.5% in the VIS, and 4.6% in the ultraviolet (UV) range. Inset: CPPB hydrogel transparency display, using two 1 mm thick regular glass ($SiO_2$) to clamp 2 mm CPPB hydrogel. **e**, Absorbance spectrum of CPPB hydrogel film measured with ATR–FTIR (Fourier transform infrared spectroscopy) spectroscopy. **f**, Thermal insulation property of CPPB hydrogel with different F127 content. **g**, Thermal insulation property of CPPB hydrogel with different water content. **h**, Thermal insulation property of CPPB hydrogel in 2 hour and temperature difference between CPPB hydrogel and control group and heat source temperature (red line). **i**, UV absorbance of CPPB and PPB.

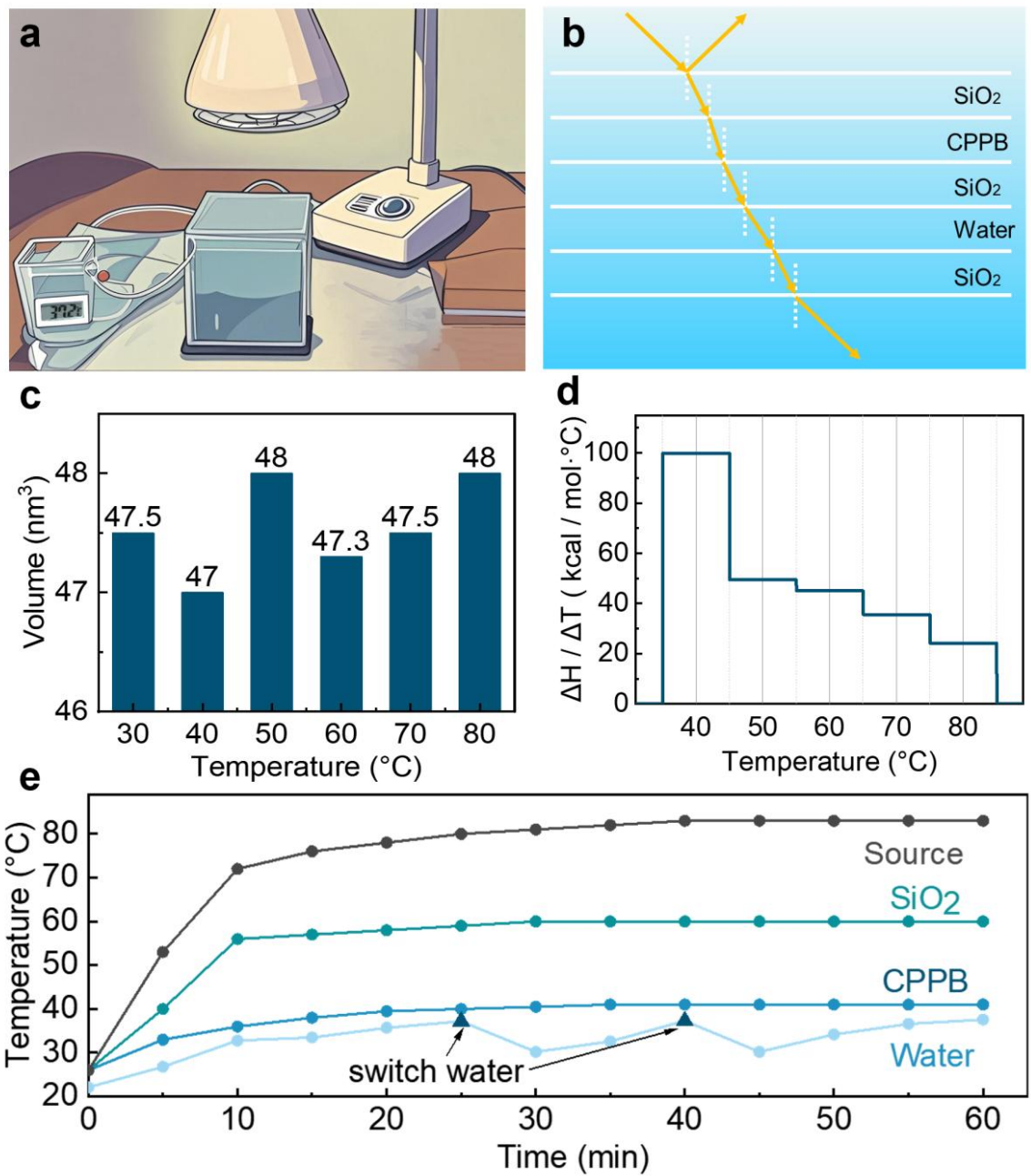

**Fig.2 Design and Properties of CPW System. a**, Physical diagram of the simplified CPW system device. **b**, Five-layer structure and refraction and reflection model of CPW system. **c**, Volume changes of CPPB hydrogel at different temperatures. **d**, Ratio of enthalpy change to temperature change for the CPPB hydrogel at different temperatures. **e**, Thermal insulation and waste heat reuse properties of the CPW system. The heat source temperature, the control glass temperature, and the temperature of CPW system. The water is changed at a temperature of 37 °C in the triangle so that waste heat can be collected as much as possible.

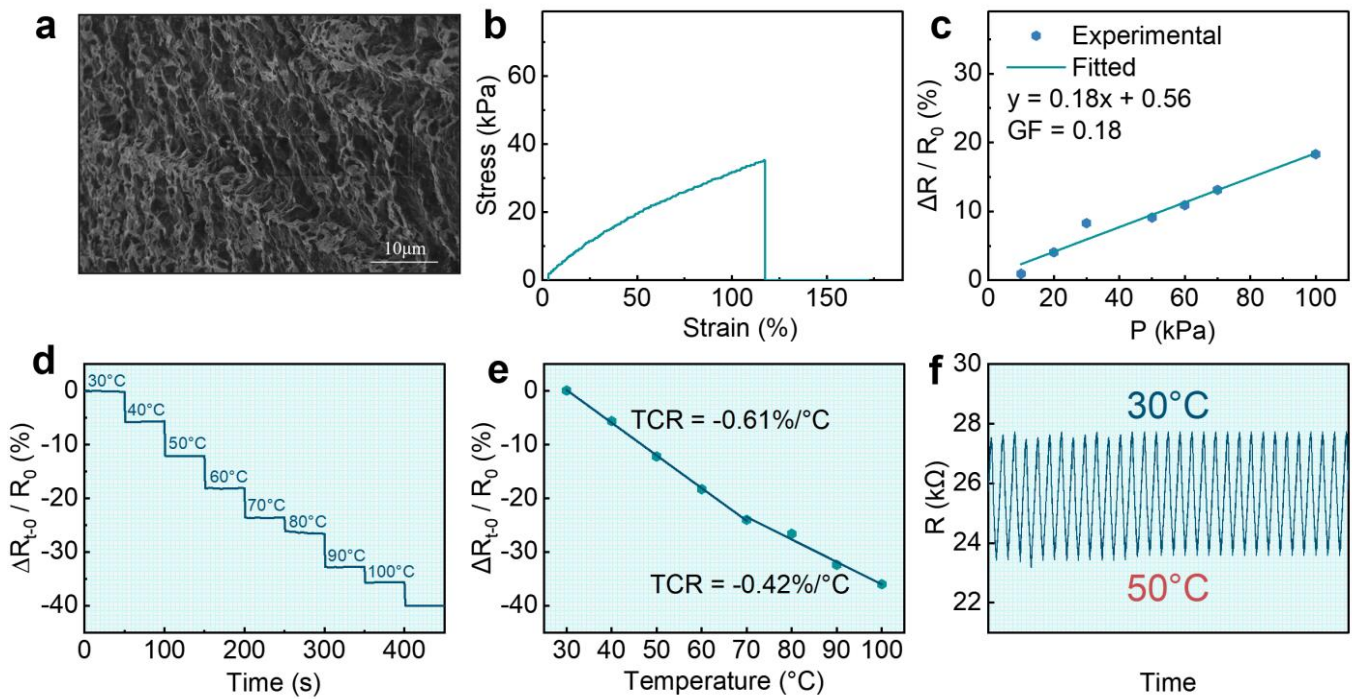

**Fig.3 Smart Sensing Integrated Design. a**, Scanning electron microscopy (SEM) of CPPB hydrogel. **b**, Stress−strain curves of CPPB hydrogel. **c**, Relative resistance changes of the CPPB hydrogel upon increasing the compressive strain from 2 to 20 kPa. **d**, Relative resistance changes of the CPPB hydrogel upon increasing the temperature from 30 to 100 °C. **e**, TCR of CPPB hydrogel. **f**, Real-time temperature responsiveness during repeated cooling and heating between 30 and 50 °C.

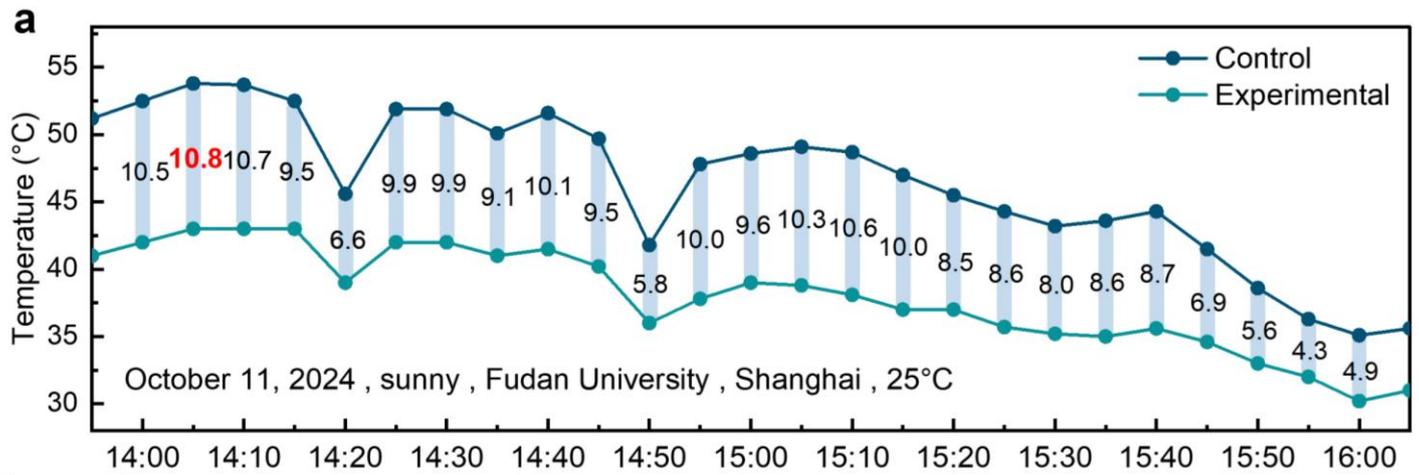
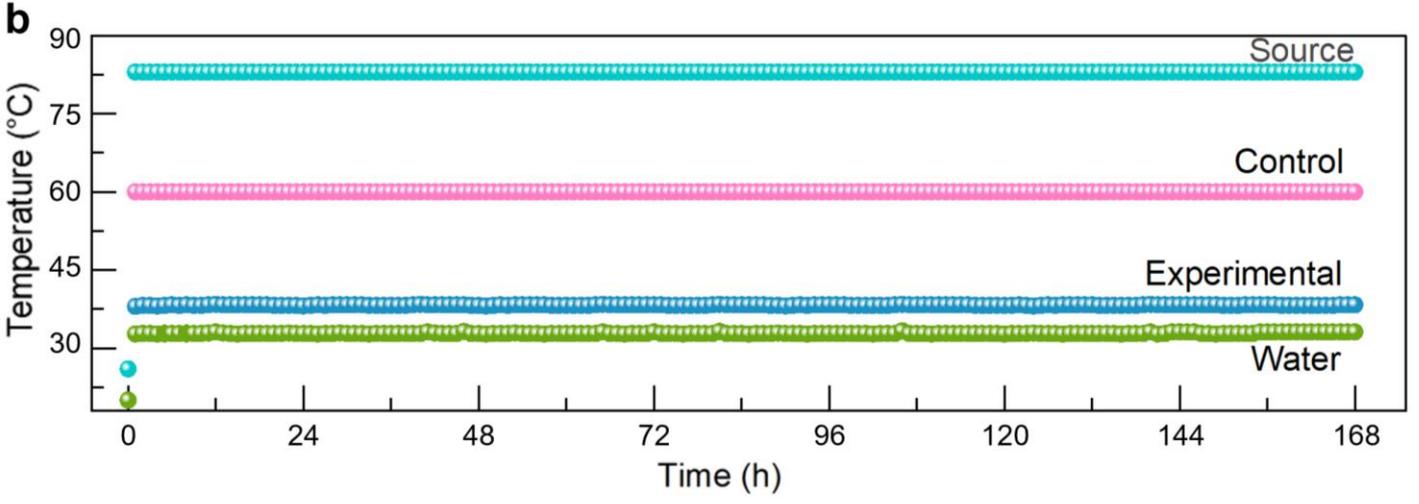
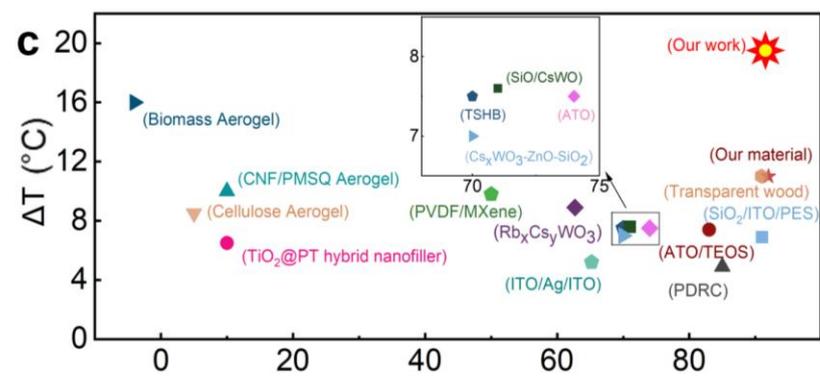
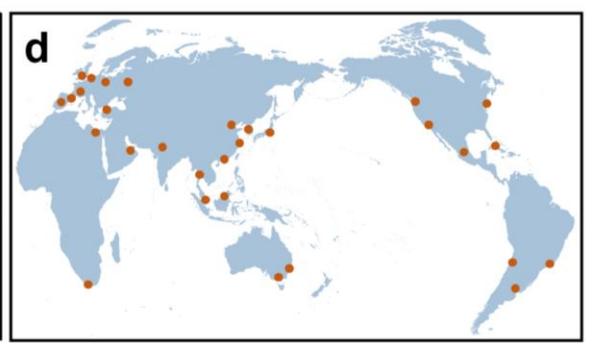
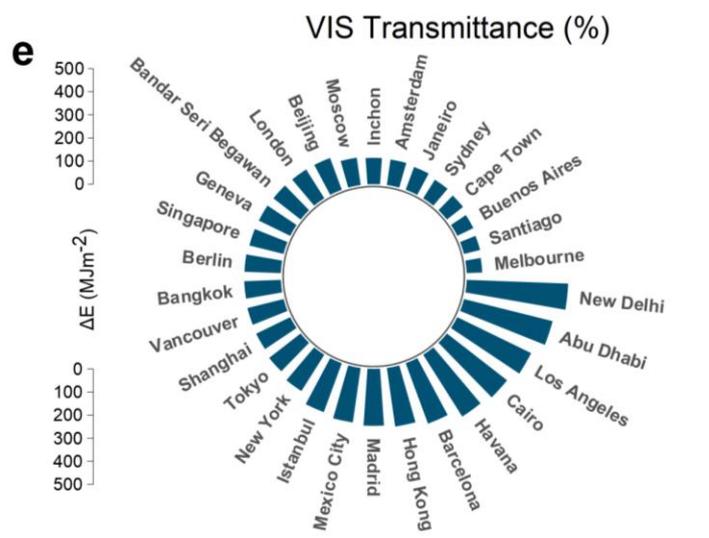
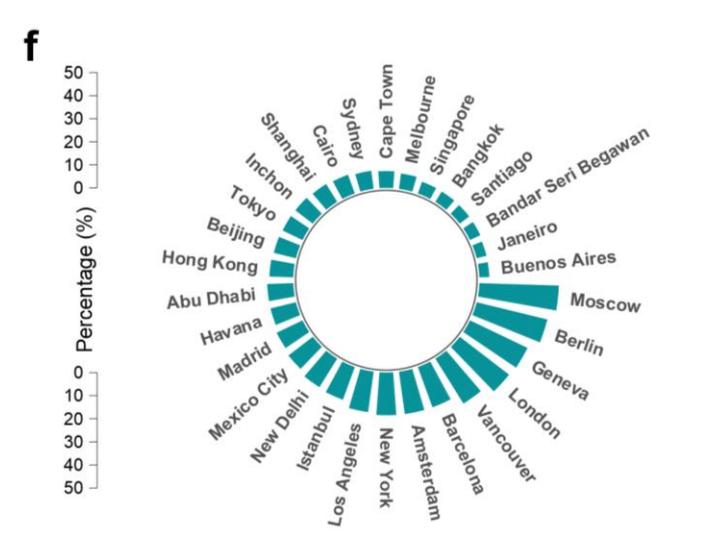

**Fig.4 Application and Energy Saving. a**, Outdoor field test on October 11, 2024, at Fudan University's Jiangwan Campus, located at 121.5°E and 31.3°N. RH = 93%. **b**, Thermal insulation of CPW system in 168 h.

**c**, Overview of the thermal insulation properties and VIS transmittance of various materials. **d**, Geographical location of 30 cities on the map. **e**, Annual cooling energy consumption of the model building using the CPW system is lower than that of the vacuum glass, based on weather data from 30 cities. **f**, Energy consumption reducing for the model building using the CPW system, based on weather data from 30 cities.